\documentstyle[aps,epsf,multicol]{revtex}
\title{2D Numerical Simulation of the Resistive Reconnection Layer.}
\author{D.~A.~Uzdensky\footnote{Currently at the University 
of Chicago.} and R.~M.~Kulsrud \\
\em{Princeton Plasma Physics Laboratory, P.O.Box 451, \\
Princeton University,  Princeton, NJ 08543}}
\date{January 7, 1999}
\begin{document}
\input psfig.sty
\maketitle
\begin{abstract}

In this paper we present a two-dimensional numerical simulation 
of a reconnection current layer in incompressible resistive 
magnetohydrodynamics with uniform resistivity in the limit 
of very large Lundquist numbers. We use realistic boundary 
conditions derived consistently from the outside magnetic 
field, and we also take into account the effect of the 
backpressure from flow into the the separatrix region. 
We find that within a few Alfv{\'e}n times the system 
reaches a steady state consistent with the Sweet--Parker 
model, even if the initial state is Petschek-like.

~
\end{abstract}

\begin{multicols}{2}

Magnetic reconnection is of great interest in many space and 
laboratory plasmas \cite{Kulsrud-1998,MRX-Yamada}, and has been 
studied extensively for more than four decades. The most important 
question is that of the reconnection rate. The process of magnetic 
reconnection, is so complex, however, that this question is still 
not completely resolved,  even within the simplest possible {\it 
canonical} model: two-dimensional (2D) incompressible resistive 
magnetohydrodynamics (MHD) with uniform resistivity $\eta$ in the 
limit of $S\rightarrow \infty$ (where $S=V_A L/\eta$ is the global 
Lundquist number, $L$ being the half-length of the reconnection layer). 
Historically, there were two drastically different estimates for the 
reconnection rate: the Sweet--Parker model \cite{Sweet-1958,Parker-1963} 
gave a rather slow reconnection rate ($E_{\rm SP} \sim S^{-1/2}$), 
while the Petschek \cite{Petschek-1964} model gave any reconnection 
rate in the range from $E_{\rm SP}$ up to the fast maximum Petschek 
rate $E_{\rm Petschek} \sim 1/\log S$. Up until the present it was 
still unclear whether Petschek-like reconnection faster than Sweet--Parker 
reconnection is possible. Biskamp's simulations \cite{Biskamp-1986} are 
very persuasive that, in resistive MHD, the rate is generally that of 
Sweet--Parker. Still, his simulations are for $S$ in the range of a 
few thousand, and his boundary conditions are somewhat tailored to 
the reconnection rate he desires, the strength of the field and the 
length of layer adjusting to yield the Sweet--Parker rate. Thus, a 
more systematic boundary layer analysis is desirable to really settle
the question.

We believe that the methods developed in the present paper
are rather universal and can be applied to a very broad
class of reconnecting systems. However, for definiteness 
and clarity we keep in mind a particular global geometry
presented in Fig.~\ref{fig-global} (although we do not use
it explicitly in our present analysis). This Figure shows
the situation somewhere in the middle of the process of 
merging of two plasma cylinders. Regions~I and II are 
ideal MHD regions: regions~I represent unreconnected flux,
and region~II represents reconnected flux. The two regions~I 
are separated by the very narrow {\it reconnection current layer}.
Plasma from regions~I enters the reconnection layer and gets 
accelerated along the layer, finally entering the {\it separatrix 
region} between regions~I and II. In general, both the reconnection 
layer and the separatrix region require resistive treatment. 

{\columnwidth=3in
\begin{figure} [tbp]
\centerline {\psfig{file=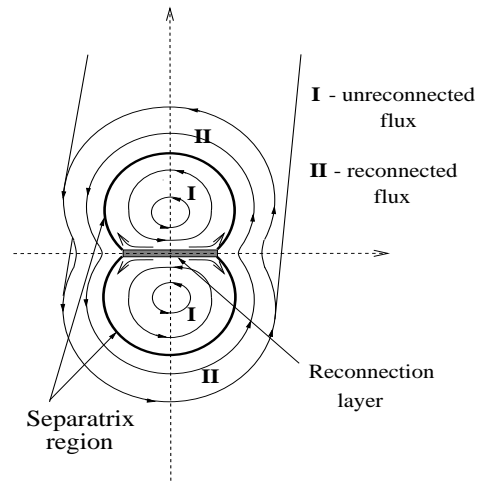,height=2.5 in,width=2.5 in,angle=90}}
\caption[The global geometry]
{The global geometry.}
\label{fig-global}
\end{figure}
}

In the limit $S\rightarrow \infty$ the reconnection rate is slow compared
with the Alfv{\'e}n time $\tau_A=L/V_A$. Then one can break the whole 
problem into the global problem and the local problem \cite{Uzdensky-1996}. 
The solution of the global problem is represented by a sequence of 
magnetostatic equilibria, while the solution of the local problem 
(concerning the narrow resistive reconnection layer and the separatrix 
region) determines the reconnection rate. The role of the global problem 
is to give the general geometry of the reconnecting system, the position
and the length of the reconnection layer and of the separatrix, and the
boundary conditions for the local problem. These boundary conditions are 
expressed in terms of the outside magnetic field~$B_{y,0}(y)$, where $y$ 
is the direction along the layer. In particular, $B_{y,0}(y)$ provides the 
characteristic global scales: the half-length of the layer~$L$, defined 
as the point where $B_{y,0}(y)$ has minimum, and the global Alfv{\'e}n 
speed, defined as $V_A=B_{y,0}(0)/\sqrt{4\pi\rho}$.

In the present paper we study the local problem using the boundary 
conditions provided by our previous analysis of the global problem 
\cite{Uzdensky-1997}. Our main goal is to determine the internal 
structure of a steady state reconnection current layer (i.e., to find the 
2D profiles of plasma velocity and magnetic field), and the reconnection 
rate represented by the (uniform) electric field~$E$. We assume 
{\it incompressible resistive MHD with uniform resistivity}. Perfect 
mirror symmetry is assumed with respect to both the $x$ and $y$~axes 
(see Fig.~\ref{fig-comp-box}). 

This physical model is described by the following three
steady state fluid equations: the incompressibility condition,
$\nabla \cdot {\bf v}=0$, the $z$~component of Ohm's law,
$\eta j_z=E+[{\bf v}\times{\bf B}]_z$, and the equation of 
motion, ${\bf v}\cdot \nabla {\bf v}=-\nabla p+ 
[j_z \hat{z}\times{\bf B}]$ (with the density set to one).

Now we take the crucial step in our analysis. 
We note that the reconnection problem is fundamentally 
a boundary layer problem, with $S^{-1}$ being the small 
parameter. This allows us to perform a {\it rescaling procedure}
\cite{Uzdensky-1998} inside the reconnection layer, to make 
rescaled resistivity  equal to unity. We rescale distances 
and fields in the $y$-direction by the corresponding global 
values ($L$, $B_{0,y}(0)$, and $V_A$), while rescaling 
distances and fields in the $x$-direction by the corresponding 
local values: $x\rightarrow x \delta_0$, $v_x\rightarrow
v_x V_A \delta_0/L$, $B_x \rightarrow B_x B_{y,0}(0) 
\delta_0/L$, $E\rightarrow E B_{y,0}(0) V_A \delta_0/L$.
Here, $\delta_0\equiv LS^{-1/2}$ is the Sweet-Parker thickness 
of the current layer. Thus, one can see that the small scale 
$\delta_0$ emerges naturally. Then, using the small parameter 
$\delta_0/L=S^{-1/2} \ll 1$, one obtains a simplified set of 
fluid equations for the rescaled dimensionless quantities: 
$$ \nabla \cdot {\bf v} = 0,                                    \eqno{(1)}  $$
$$ E = {{\partial B_y}\over{\partial x}} - v_x B_y + v_y B_x,   \eqno{(2)}  $$
(where the first term on the right hand side (RHS) is the resistive term) 
and
$$ {\bf v}\cdot \nabla v_y = -{{\partial p}\over{\partial y}} + 
B_x {{\partial B_y}\over{\partial x}}.                          \eqno{(3)}  $$

In the last equation (representing the equation of motion in the 
$y$-direction, along the current layer) the pressure term can be 
expressed in terms of $B_y(x,y)$ and the outside field $B_{0,y}(y)$ 
by using the vertical pressure balance (representing the $x$-component 
of the equation of motion, across the current layer):
$$ p(x,y) = {{B_{y,0}^2(y)}\over 2} - {{B_y^2(x,y)}\over 2}.    \eqno{(4)} $$

We believe that this rescaling procedure captures all the important dynamical 
features of the reconnection process.

The problem is essentially two-dimensional, and requires a 
numerical approach. Therefore, we developed a numerical code
for the main reconnection layer, supplemented by another code 
for the separatrix region. The solution in the separatrix region
is needed to provide the downstream boundary conditions for the 
main layer (see below).

The steady state was achieved by following the true time evolution of the 
system starting with initial conditions discussed below. The time
evolution was governed by two dynamical equations:
$$ \dot{\Psi} = -\nabla\cdot({\bf v} \Psi) + 
{{\partial^2 \Psi}\over{\partial x^2}} + 
\left( \eta_y{{\partial^2 \Psi}\over{\partial y^2}} \right),     \eqno{(5)} $$
$$ \dot{v_y} =  -\nabla\cdot({\bf v} v_y) - 
{d\over{dy}} \left[{{B_{y,0}^2(y)}\over 2}\right] +
\nabla \cdot ({\bf B} B_y) +
\left( \nu_y{{\partial^2 v_y}\over{\partial y^2}} \right).       \eqno{(6)} $$

(Small artificial resistivity $\eta_y$ and viscosity $\nu_y$
were added for numerical stability.) The natural unit of time 
is the Alfv{\'e}n time $\tau_A = L/V_A$. The magnetic flux 
function~$\Psi$ is related to ${\bf B}$ via $B_x=-\Psi_y$, 
and $B_y=\Psi_x$. At each time step, $v_x$ was obtained by 
integrating the incompressibility condition: $v_x(x,y) = 
-\int_0^x (\partial v_y/\partial y) dx$. Note that this means 
that we do not prescribe the incoming velocity, and 
hence the reconnection rate: the system itself determines how
fast it wants to reconnect. 

We used the finite difference method with centered derivatives
in $x$ and $y$ (second order accuracy). The time derivatives
were one-sided. The numerical scheme was explicit in the $y$
direction. In the $x$ direction the resistive term $\partial^2
\Psi /\partial x^2$ was treated implicitly, while all other terms
were treated explicitly. Calculations were carried out on a rectangular 
uniform grid. We considered only one quadrant because of symmetry 
(see Fig.~\ref{fig-comp-box}). More details can be found in 
Ref.~\cite{Thesis}.

{\columnwidth=3in
\begin{figure} [tbp]
\centerline {\psfig{file=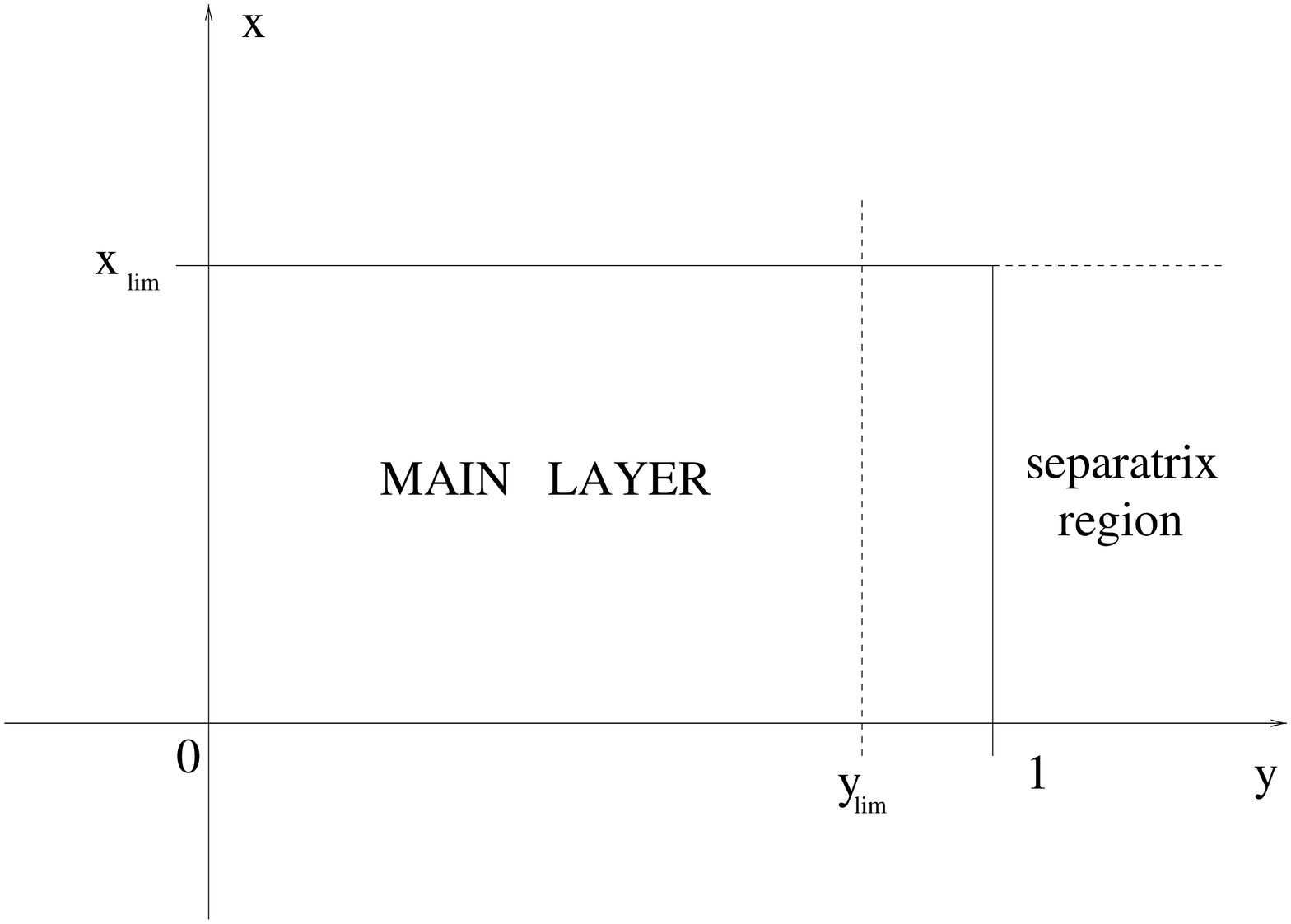,height=2.0 in,width=3 in}}
\caption[Computational box used in the numerical simulation]
{Computational box used in the numerical simulation.}
\label{fig-comp-box}
\end{figure}
}

The boundary conditions on the lower and left boundaries 
were those of symmetry (see Fig.~\ref{fig-comp-box}). On 
the upper (inflow) boundary $x=x_{\rm lim}$ the boundary 
conditions were $\partial v_y/ \partial x =0$ (which worked 
better than $v_y=0$) and $B_y(x_{\rm lim},y)=B_{0,y}(y)$ --- 
the prescribed outside magnetic field. In our simulations 
we chose $B_{0,y}(y) = B_0+(1-B_0)\sqrt{1-y^2}$ with $B_0=0.3$, 
consistent with the global analysis of our previous paper 
\cite{Uzdensky-1997}.

The boundary conditions on the right (downstream) boundary
cannot be given in a simple closed form. Instead, they require 
matching with the solution in the separatrix region, which 
itself is just as complicated as the main layer. Therefore,
we have developed a supplemental numerical procedure for the
separatrix region. Noticing that in the separatrix region 
the resistive term should not qualitatively change the solution, 
we adopt a simplified ideal-MHD model for the separatrix. This 
model is expected to give a qualitatively correct picture of the
dynamical influence of the separatrix region on the main layer, 
and thus a sufficiently reasonable downstream boundary conditions 
for the main layer. In particular, our model includes the effects 
of the backpressure that the separatrix exerts on the main layer.

The advantages of our approach are: ({\it i}) use of the rescaled 
equations takes us directly into the realm of $S\rightarrow \infty$;
({\it ii}) we do not prescribe the incoming velocity $v_x(x_{\rm lim},y)$ 
as a boundary condition: $v_x$ is determined not by the $x$-component 
of the equation of motion, but rather by $v_y$ via the incompressibility 
condition. As a result, we do not prescribe the reconnection rate; 
({\it iii}) the use of true time evolution guarantees that the 
achieved steady state is two-dimensionally stable; ({\it iv}) 
we have a realistic variation of the outside magnetic field 
along the layer, with the endpoint~$L$ of the layer clearly 
defined as the point where $B_{0,y}(y)$ has minimum (see Ref.~\cite
{Uzdensky-1997}).

Let us now discuss the {\it results} of our simulations. 
We find that, after a period of a few Alfv{\'e}n times, {\it 
the system reaches a Sweet--Parker-like steady state, 
independent of the initial configuration}. In particular, 
when we start with a {\it Petschek-like} initial conditions 
(see Fig.~\ref{j-3d}a), the high velocity flow rapidly sweeps 
away the transverse magnetic field $B_x$ (see Fig.~\ref{Bx(0,y)-t-petschek}). 
This is important, because, for a Petschek-like configuration to exist,
the transverse component of the magnetic field on the midplane, $B_x(0,y)$, 
must be large enough to be able to sustain the Petschek shocks in the field 
reversal region. For this to happen, $B_x(0,y)$ has to rise rapidly 
with~$y$ inside a very short diffusion region, $y < y_* \ll L$ (in 
the case $E_{\rm init} = 2E_{\rm SP}$, presented in Fig.~\ref{j-3d}a, 
$y_*=L/4$), to reach a certain large value ($B_x=2$ for~$E_{\rm init}= 
2E_{\rm SP}$) for $y_* < y < L$. While the transverse magnetic flux 
is being swept away by the plasma flow, it is being regenerated by the 
merging of the $B_y$~field, but only at a certain rate and only on a global 
scale in the $y$-direction, related to the nonuniformity of the outside 
magnetic field~$B_{y,0}(y)$, as discussed by Kulsrud \cite{Kulsrud-1998}. 
As a result, the initial Petschek-like structure is destroyed, and the 
inflow of the magnetic flux through the upper boundary drops in a 
fraction of one Alfv{\'e}n time. Then, after a transient period, 
the system reaches a steady state consistent with the Sweet--Parker 
model.

{\columnwidth=3in
\begin{figure} [t]
\centerline
{\psfig{file=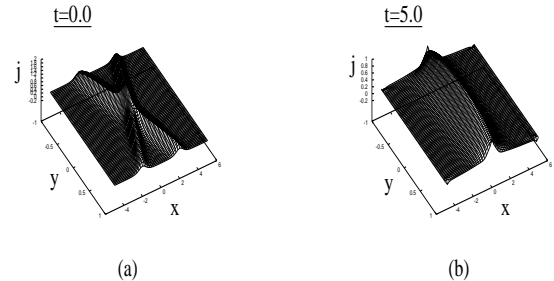,height=1.5in,width=3in}}
\vspace{10pt}
\caption{The current density $j(x,y)$: {\bf (a)} at $t=0$ for 
Petschek-like initial conditions with $E_{\rm init}=2E_{\rm SP}$,
and {\bf (b)} in the final steady state (at $t=5$), which 
corresponds to the Sweet--Parker solution. (All four quadrants 
are shown for clarity).}
\label{j-3d}
\end{figure}
}

{\columnwidth=3in
\begin{figure} [tbp]
\centerline 
{\psfig{file=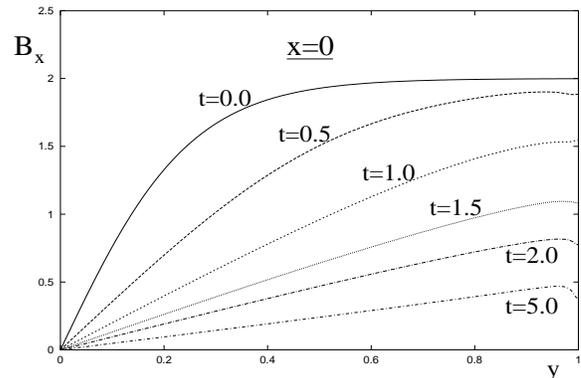,height=2.0in,width=3in}}
\caption[The time evolution of the variation of the transverse magnetic 
field $B_x(0,y)$ along the midplane $x=0$ for the Petschek-like initial 
conditions]
{The time evolution of the variation of the transverse magnetic field 
$B_x(0,y)$ along the midplane $x=0$ for the Petschek-like initial conditions.}
\label{Bx(0,y)-t-petschek}
\end{figure}
}

We believe that the fact that we rescaled $x$ using the 
Sweet--Parker scaling does not mean that we prescribe the 
Sweet--Parker reconnection rate. Indeed, if the reconnecting 
system wanted to evolve towards Petschek's fast reconnection, 
it would then try to develop some new characteristic structures, 
e.g. Petschek-like shocks, which we would be able to see. Note 
that, if Petschek is correct, then there should be a range of 
reconnection rates including those equal to any finite factor 
greater than one times the Sweet--Parker rate~$E_{\rm SP}$. 
However, in our simulations we have demonstrated that there 
is only one stable solution and that it corresponds to~$E=E_{\rm SP}$.
In this sense we have demonstrated that Petschek 
must be wrong since reconnection can not even go 
a factor of two faster than Sweet--Parker, let 
alone almost the entire factor of~$\sqrt{S}$. 
There seems no alternative to the conclusion 
that fast reconnection is impossible. 

It is interesting that in Petschek's original paper 
the length of the central diffusion region $y_*$ is an
undetermined parameter, and the reconnection velocity 
$v_{\rm rec}$ depends on this parameter as $V_A (L/y_*)^2 /
\sqrt{S} $. If $y_*$ is taken as small as possible then 
Petschek finds that $v_{\rm rec} \sim V_A/ \log(S)$.
However, $y_*$ should be determined instead by balancing 
the generation of the transverse field $B_x$ against its 
loss by the Alfv{\'e}nic flow (it should be remarked that
Petschek did not discuss the origin of this transverse
field in his paper). As we discussed above, this balance 
yields $y_* \approx L$, with the resulting unique rate 
equal to that of Sweet--Parker. This results are borne 
out by our time dependent numerical simulations.

The final steady state solution is represented in 
Fig.~\ref{j-3d}b. It corresponds to $x_{\rm lim}=5.0$, 
$y_{\rm lim}=1.0$, $\eta_y=\nu_y=0.01$. 
We see that the solution is consistent with the Sweet--Parker
picture of reconnection layer: the plasma parameters change
on the scale of order $\delta_0$ in the $x$ direction
and on a global scale $L$ in the $y$-direction.
The reconnection rate in the steady state is surprisingly close
to the typical Sweet--Parker reconnection rate $E_{\rm SP}=\eta^{1/2}
V_A B_{y,0}(0)$. The solution is numerically robust: it does 
not depend on $x_{\rm lim}$, $y_{\rm lim}$ or on the small 
artificial resistivity~$\eta_y$ and viscosity~$\nu_y$.

Several things should be noted about this solution.
First, $j(x,y)\rightarrow 0$ (and $B_y(x,y) \rightarrow B_{0,y}(y)$) 
monotonically as $x\rightarrow \infty$, meaning that there is no flux 
pile-up. Second, as can be seen from Fig.~\ref{Bx(0,y)-t-petschek}, 
$B_x(x=0,y) \sim y$ near $y=0$, contrary to the cubic behavior 
predicted by Priest--Cowley \cite{Priest-Cowley-1975}. This is 
due to the viscous boundary layer near the midplane $x=0$ and 
the resulting nonanalytic behavior in the limit of zero viscosity, 
as explained in Ref.~\cite{Uzdensky-1998}. Third, there is a sharp 
change in $B_x$ and $j$ near the downstream boundary $y=y_{\rm lim}=1$, 
due to the fact that in the separatrix region we neglect the resistive 
term (which is in fact finite). 

It appears that the destruction of the initially-set-up Petschek-like 
configuration and its conversion into a Sweet-Parker layer happens so 
fast that it is determined by the dynamics in the main layer itself 
and by its interaction with the upstream boundary conditions (scale 
of nonuniformity of $B_{0,y}$), as outlined above. Therefore, the fact 
that our model for the separatrix does not describe flow in the separatrix 
completely accurately seems to be unimportant. However, for the solution 
of the problem to be really complete, a better job has to be done in 
describing the separatrix dynamics, and, particularly, the dynamics
in the very near vicinity of the endpoint of the reconnection layer. 
A proper consideration of the endpoint can not be done in our rescaled 
variables, and a further rescaling of variables and matching is needed.

To summarize, in this paper we present a definite solution to a particular
clear-cut, mathematically consistent problem concerning the internal
structure of the reconnection layer within the canonical framework 
(incompressible 2D MHD with uniform resistivity) with the outside 
field $B_{0,y}(y)$ varying on the global scale along the layer.
Petschek-like solutions are found to be unstable, and the system 
quickly evolves from them to the unique stable solution corresponding 
to the Sweet--Parker layer. The reconnection rate is equal to
the (rather slow) Sweet--Parker reconnection rate, $E_{\rm SP}
\sim 1/\sqrt{S}$. This main result is consistent with the results 
of simulations by Biskamp \cite{Biskamp-1986} and also with the 
experimental results in the MRX experiment \cite{MRX-Ji}.

Finally, because the Sweet--Parker model with classical (Spitzer)
resistivity is too slow to explain solar flares, one has to add 
new physics to the model, e.g., locally enhanced anomalous 
resistivity. This should change the situation dramatically, and may 
even create a situation where a Petschek-like structure with fast 
reconnection is possible (see, for example, Refs.~\cite
{Ugai-Tsuda-1977,Scholer-1989,Kulsrud-1998}).

We are grateful to D.~Biskamp, S.~Cowley, T.~Forbes, M.~Meneguzzi, 
S.~Jardin, M.~Yamada, H.~Ji, S.~Boldyrev, and A.~Schekochihin for 
several fruitful discussions. This work was supported by Charlotte 
Elizabeth Procter Fellowship, by the Department of Energy Contract 
No. DE-AC02-76-CHO-3073, and by NASA's Astrophysical Program under 
Grant NAGW2419.

\end{multicols}

\end{document}